%Paper: hep-th/9408153
%From: Lev Vaidman <vaidman@ccsg.tau.ac.il>
%Date: Sun, 28 Aug 1994 17:16:54 +0300 (IDT)

\magnification 1200
\baselineskip 22pt

\centerline{\bf The Meaning of Protective Measurements}

\centerline{Yakir Aharonov$^{a,b}$, Jeeva Anandan$^b$ and Lev Vaidman$^a$}

\vskip 1cm

\centerline{\it $^a$  School of Physics and Astronomy}
\centerline{\it Raymond and Beverly Sackler Faculty of Exact Sciences}
\centerline{\it Tel-Aviv University, Tel-Aviv, 69978 ISRAEL}

\vskip .5cm

\centerline{\it $^b$ Physics Department, University of South Carolina}
\centerline{\it Columbia, South Carolina 29208, U.S.A.}

\vskip 1.5cm
\centerline{ Abstract}
Protective measurements, which we have introduced recently, allow to measure
properties of the  state of a single quantum system and
even  the Schr\"odinger wave itself. These measurements require a
protection, sometimes due to an additional procedure and sometimes due to the
potential of the system itself. The analysis of the protective
measurements is presented and it is argued, contrary to  recent claims, that
they measure  the quantum state
  and not the protective potential.
Some other misunderstandings concerning our proposal are clarified.
 \vskip 2cm

\break

In this work we will analyze the meaning of our recent proposal of
``protective measurements'' which allow measuring the Schr\"odinger wave of
a single particle [1,2]. There were several discussions of our proposal some
of which challenging the validity of our results [3-7]. We shall briefly
repeat our argument and will answer the questions raised by the critics.
We, ourselves, believe that the objection to our proposal which is most
difficult to refute is the one which was not raised by the critics. Since
our measurement requires a long time, it is natural to think that what is
measured is not the property of the Schr\"odinger wave at a given time, but
the time average during the period of the measurements. In Refs. (1) and (2) we
make a
detailed analysis of this question and  we will not discuss it here.

 At present, the commonly accepted interpretation of the Schr\"odinger wave is
due to Born.  He proposed to interpret the wave intensity not as the
density of distribution of actual matter, as Schr\"odinger first imagined, but
as
a probability density for the presence of a particle. Schr\"odinger,
however, wanted to believe that his wave represents a single particle: the
wave is an extended object really moving in space.   Born's
interpretation was supported by the fact that nobody knew how to measure the
density of the Schr\"odinger wave on a single system. There was a general
belief that the Schr\"odinger wave can only be tested for an ensemble of
particles. We have proposed  a new type of measurements: ``protective
measurements'' which
allow direct measurement of the  Schr\"odinger wave density on a
single particle. We have shown that one  can
simultaneously measure the density and the current of the Schr\"odinger
wave in many locations. The results of these measurements then allow to
reconstruct (in an arbitrary chosen gauge) the Schr\"odinger wave.

In order to analyze the meaning of our proposal we will start the
discussion not by considering a measurement of the whole Schr\"odinger
wave, but of a single property of the quantum state: an expectation value
of an operator.  The essential novel point is that we can measure
an expectation value of a quantum variable for a state which is not the
eigenstate of the corresponding operator even when performing measurements just
on a
single system; or, what is even more important, that we obtain the
expectation value directly for a single system and not as a statistical average
of
eigenvalues for an ensemble. In the standard interpretation the expectation
value is defined
as a certain statistical function of the eigenvalues. Therefore, since we can
obtain the expectation value without knowing the eigenvalues, we can give a
new interpretation.

In the standard interpretation the expectation values of
variables are not considered as physical properties of a single system
because in  the outcome of the standard measuring procedure  only  one of
the eigenvalues is observed. If the system, prior to the
measurement of a variable $A$, is not in an eigenstate of $A$, then its
quantum state is invariably changed due to the standard measuring
procedure. A natural idea to prevent this change is to weaken the coupling
with the measuring device. Indeed, in this case the state is not changed
significantly, but then the pointer of the measuring device hardly
moves. Its shift due to the measurement is smaller than its uncertainty,
and therefore we cannot get information from the measurement.  To remedy
this we can increase the time of the coupling between the system and the
measuring device. Then, if the state is constant during the measurement,
the velocity of the pointer variable will also be constant and the total
shift, which is proportional to duration of the interaction, will be large
enough.  However, under normal circumstances the state of the system is
not constant during the measurement. Weak coupling leads to a small rate of
change of the state. However, in order  to obtain a distinguishable shift
it  requires a long time, and therefore, we invariably change the
state. Thus, the reading of the measuring device will not correspond to the
quantum state which the system had prior to the measurement, but to some
time average depending on the evolution of the quantum state influenced by
the measuring procedure. Therefore, in order to measure the quantum state,
or a property of the quantum state such as expectation value of a variable
(when the state is not an eigenstate of the corresponding operator) we
need, in addition to the standard weak and long measuring interaction, a
procedure which will protect the state from changing during the measuring
interaction.

The simplest protection procedure is introducing a protective potential  such
that the
quantum state of the system will be a nondegenerate eigenstate of the
Hamiltonian. In fact, in many important cases this protection is given by
nature: almost isolated systems  will eventually decay to their  ground state
or to  some stable excited
state.
 In order to protect the state the protection potential need  not
 act all the time. It is possible to switch it on for frequent
very short periods of time such that most of the time the system evolves
under the original Hamiltonian. The other proposed  protection scheme -
frequent testing that the state have
not changed - is similar to this method.

As an example of a simple protective measurement, let us consider a particle in
a discrete nondegenerate energy
eigenstate $\Psi (x)$. The standard von Neumann
procedure for measuring the value of an observable $A$ in
this state involves an
 interaction Hamiltonian
$$
 H = g(t) P A,\eqno(1)
$$
coupling the system to a measuring device, or pointer, with coordinate
and momentum denoted, respectively, by $Q$ and $P$.  The time-dependent
coupling $g(t)$ is normalized to $\int g(t) dt =1$ and  the initial state
of the pointer is taken to be a Gaussian centered around zero.

In standard impulsive measurements, $g(t) \neq 0 $  for only a very
short time interval.  Thus, the interaction term dominates the rest of
the Hamiltonian, and the time evolution
$
e^{-{i\over \hbar}P A}
$
leads to a correlated state: eigenstates of $A$ with eigenvalues $a_n$
are correlated to measuring device states in which the pointer is
shifted by these values $a_n$.  By contrast, the protective
measurements of interest here utilize the opposite limit of extremely
slow measurement.  We take $g(t) = 1/T$ for most of the time $T$ and
assume that $g(t)$ goes to zero gradually before and after the period
$T$. We choose the initial state of the measuring device  such
that the canonical conjugate $P$ (of the pointer variable $Q$) is
bounded.  We also assume that $P$ is a constant of motion not only of the
interaction Hamiltonian (1), but of the whole Hamiltonian. For $g(t)$ smooth
enough we obtain an adiabatic process in
which the particle cannot make a transition from one energy eigenstate
to another, and, in the limit $T \rightarrow \infty$, the interaction
Hamiltonian does not change the energy eigenstate.  For any given value of
$P$, the energy of the eigenstate shifts by an infinitesimal amount
given by first order perturbation theory:
$$\delta E = \langle H_{int}  \rangle  = {{
\langle
A\rangle P}\over T}.
\eqno(2)$$
The corresponding time evolution $ e^{-i P \langle A\rangle /\hbar} $ shifts
the
pointer by the average value $\langle A
\rangle $.  This  result contrasts with the usual (strong)
measurement in which the pointer shifts by one of the eigenvalues of
$A$.
  By measuring the averages of a sufficiently large number of
variables $A_n$, the full Schr\"odinger wave $\Psi (x)$ can be
reconstructed to any desired precision.

The adiabatic measurements described above can be performed on a single
system only if the measured quantum state is protected.
However, when the mathematical
expression of protected and unprotected states are identical, the quantum
system behaves identically for all standard (impulsive and strong) quantum
measurements. Indeed,
the probability for various eigenvalues are the same for protected and
unprotected states. Here we assume that the strength of the impulsive
measurements (which is infinite in  ideal measurements) is much bigger
than the strength of the protection procedure. Therefore, the state being
protectively observed may be regarded as the same state if it were unprotected.

The identical probabilities can be tested only on ensembles of identically
prepared systems. Thus, to compare protected and unprotected states we have
to consider an ensemble of identical protected states and an ensemble of
identical (and the same) unprotected states. It is interesting that these
ensembles are
identical not only for ideal measurements, but also for {\it weak
measurements} [8]. If we have large ensemble of quantum systems in the
same state, then the protection procedure is not necessary. We can have
weak short coupling between the measuring device and all systems of the
ensemble. This
coupling does not change significantly the quantum state, but  the
pointer moves a distance much larger than its uncertainty due to the
combined effect of all the systems of the ensemble, showing the
expectation value of measured variable. Although we use here an ensemble,
the measurement is very different from the standard procedure: in no stage
of the measurement we obtain the eigenvalues of the measured variable. Each
system  in the ensemble contributes the shift of the pointer proportional to
the expectation
value (and not to one of the eigenvalues). It is interesting to note that for a
large enough ensemble we can tune the coupling to the measuring device
such that we will get a reliable outcome of the measurement of the
expectation value of the measured variable practically without
changing  the states of the systems: if we will test all systems after our
measurement, then the probability to find even one system  not in its
original state can be made arbitrary small.

The main point of the majority of works criticizing our proposal was  that we
cannot measure an  unknown quantum state of a single
system since  we cannot protect unknown quantum states. But we  never claimed
otherwise. More than this, we have claimed the opposite. If there was a
procedure
which allows to measure an
unprotected unknown quantum state, then it is possible to  distinguish between
nonorthogonal states. The latter, however, contradicts  unitarity of
quantum theory: the scalar product
between branches corresponding to these two states changes from nonzero to
zero when the measurement is completed. The answer is that there is no
universal protection for all states. The nonorthogonal states require
different protections; if we do not know the state we cannot protect it.
 What
we have shown is that two nonorthogonal {\it protected} states
can be distinguished. The unitarity paradox disappears since the quantum
state of the environment  in the two cases are orthogonal before the
measurement [2]. Indeed, the states of the protection device for protection of
two
different  nonorthogonal states must be orthogonal. Unprotected
unknown nonorthogonal states cannot be distinguished.

We want to clarify the issue of the collapse of the Schr\"odinger wave in our
procedure.
 Our adiabatic measurement of
an expectation value of an operator
performed on a protected state does not cause collapse. But, if the system is
in a superposition of several quantum  protected  states (i.e., in the
superposition of the eigenstates of the protection potential), then the
state is not protected. Our measuring procedure will cause a collapse to
one of the protected states. Indeed, the coupling to the measuring device
will cause a correlation between the protected states and the states of
the pointer of the measuring device. Then,  in case that the expectation
value for that eigenstate
differs from that for the other eigenstates of the superposition, the
process of macroscopical reading of the pointer variable will result in the
collapse of the quantum state. Another collapse of the quantum state related to
our proposal
occurs if the initial state is unknown and we switch on the protection for a
certain state (such as the
strong magnetic field in the example of the spin measurement).   So we have
not ``solved'' the problem of the collapse of the state in quantum
measurement, but we have never claimed otherwise.

 We cannot measure an unprotected state. We cannot protect an unknown state.
So,
it seems that to measure the state we have to know it first, but then what
is the purpose of the measurement? We believe that even if we measure a
known property, it is still a measurement. Our measuring procedure is not
some sophisticated method of reading known information. We use a coupling to
the measured variable and look directly on the reading of the pointer of
the measuring device. We use the  information  about the quantum
state prior to the adiabatic coupling  to fix
the strength of the measuring coupling. In fact, this aspect is common to
all  physical measurements: we need to have some prior
information before the measurement in order to choose an appropriate
measuring apparatus. Although it is clear that in our method of the measuring
of the expectation
value the outcome is {\it measured} and not calculated from known quantum
state, to make a decisive argument we can show that in our procedure we do
obtain information which was not known before. To this end let us split our
procedure into two stages. The first is a protection, made by one
experimenter or even just by nature, and the second is the weak measuring
procedure which yields the expectation value. The second stage is performed
by another experimenter who does not know the state. She only knows that the
state is protected and what is the degree of protection. If the protection
is due to the energy conservation, i.e., the state is a nondegenerate energy
eigenstate, then the only information which is needed is a lower bound of the
energy gaps to other levels.  This is enough to fix the strength and the
adiabaticity of the  measuring coupling. Thus, the expectation value of any
operator
can be found, and even several such measurements can be performed, which
together will yield the quantum state of the system. The precision of the
measurement of the Schr\"odinger wave can be improved to any given limit by
increasing
the time of the interaction. What persuades us that the outcome of our
procedure is the property of our single system is that it does not
depend on the particular form of the measuring interaction. {\it Any}
adiabatic measurement of a given observable performed on quantum system in
protected state (with {\it any} valid protection mechanism) will invariably
yield the same
outcome.  In order to demonstrate the  new information we can obtain, consider
an
example of a system in an unknown potential for which we know approximately
part of the spectrum of energies and we know, by leaving the system isolated
for a long time, that the system has decayed to the  ground state. Assume that
we  have just one
such system and nobody knows the exact form of the potential. In this case
we can perform adiabatic measurements of a set of variables which will
yield the quantum state of the system. All this by measuring just one
system and without obtaining the eigenvalues of the measured variables.

Another common line of the critique of our proposal was the claim that  we are
measuring  the protection potential and not the quantum state of the
system. Indeed, our  measurement on a single system succeeds only if there
is a protection, so it is natural to believe that it is the protection that
has been measured. The  example of a measurement of a spin-1/2  particle
invites
to accept this interpretation because of accidental one to one correspondence
between protection (strong magnetic field) and the state (the spin
polaraized parallel to the magnetic field). On the other hand, in the
example of a particle bound in a certain potential, there is no simple
connection between the potential and the state. In order to find the
density of the Schr\"odinger wave in a certain region, without protective
measurements, one has to perform
elaborate {\it calculations} based on the information about the potential
everywhere.
The density of the Schr\"odinger wave can manifest itself in numerous
adiabatic measurements. We can make coupling to any kind of  charge
in any (although necessarily adiabatic and weak) form. All such measurements
will yield the
density of the wave. These measurements will yield the same results for
infinite number of various protective potentials which are all characterized by
having a nondegenerate eigenstate with this wave density in the chosen
location. These persuade us that our procedure is the measurement of the
density of the Schr\"odinger wave.

An even stronger argument against the interpretation of our procedure as a
measurement of the protection potential rather than the protected state  is
that in general there are many
possible different protection procedures which can protect the same state.
A general form of the protection Hamiltonian for a state $|\Psi_0 \rangle$ is
 $$
 H = G_0 (1 - |\Psi_0\rangle \langle \Psi_0|) + \sum G_i
|\Psi_i\rangle \langle \Psi_i|, \eqno(3)
$$
 where the only requirements on the terms in the sum are: $\langle \Psi_0
|\Psi_i \rangle = 0 $ for all $i$, and the energies
in the spectrum of the states $|\Psi_i \rangle$  are far from the energy of the
state $|\Psi_0\rangle$.
Thus, our measurement cannot be considered as a measurement of a protection
Hamiltonian, but it  can be considered as a  measurement of a
certain property of the protection Hamiltonian, i.e. the property to have as
an eigenstate this quantum state characterized by the  expectation values of
various operators.  We
measure the property of the protection which is characterized completely by
the quantum state, so it is just a matter of taste to call or not
this property the quantum state.

The simplest protection Hamiltonian $ H = G_0 (1 - |\Psi_0\rangle \langle
 \Psi_0|)$ can be immediately generalized also for the time-dependent
 quantum state $|\Psi_0 (t)\rangle$ : $$ H (t) = G_0 (1 - |\Psi_0
 (t)\rangle \langle \Psi_0 (t)|) . \eqno(4) $$ However, these protection
 Hamiltonians are, in general, nonlocal.  In our discussion, in the
 framework of the nonrelativistic quantum theory, we can consider any
 Hamiltonian. Still it is an interesting question whether we can find a
physical
 protection procedure for nonlocal states. A local potential $V(r)$ can give
 a protection to all its nondegenerate eigenstates.  Note that for a
 Hamiltonian $H = p^2/2m +V(r)$ the energy eigenstate essentially defines
 the potential, so there is strong correspondence between the protected
 state and protection Hamiltonian. Also, local potential cannot protect the
 state which is nonvanishing on a few disconnected regions. To protect these
 states, and to restore the freedom of choosing various protections to the
 same state, we can use the following, rather artificial, but conceptually
 important method.  As we mentioned before, the protection Hamiltonian does
 not have to act all the time. We can take the protection Hamiltonian which
 acts only for small but frequent periods of time. Then the procedure is to
 quickly bring   all parts of the state  to a single location, then switch on
 strong short protection Hamiltonian, then bring the state back and leave
it for a period
 of time without protection Hamiltonian. The periods of time without
 protection interaction can be made much longer than the periods of time of
 the protection procedure. Then the disturbance due to these times of our
 measuring device can be neglected and the outcome of the adiabatic
 coupling to any variable of the system will be its expectation value in
 the protected state.

There is a position in which the only reality in quantum theory is a set of
outcomes of measurements, and it is not allowed to discuss the reality of a
system between the measurements. It is not that this attitude is incorrect
-- it is absolutely consistent, but we believe that in this minimalistic
approach we lose a lot of important physics.  So we have to look for
possible candidates for ``reality''.  Unruh [4], following the standard
approach, considers the dynamical variables as immediate natural candidates
for the ontology of quantum theory. But as he himself points out, the
dynamical variables frequently do not have certain values (which allowed to
be only some eigenvalues). For example, a spin-1/2 particle in an EPR-Bohm
pair has no definite spin value for spin components in any direction.
Thus, the choice is, either to accept that it has no ``spin'' reality at
all, or to accept that it has many different realities simultaneously.  The
quantum state however, does not have  this difficulty: the Schr\"odinger wave
is
unique. Together with quantum state, the expectation values of all
dynamical variables are well defined and we can perform a direct
measurement of the expectation values using our adiabatic measuring
procedure. Therefore, the quantum state and the expectation values of
dynamical variables and not the eigenvalues are the entities which can be
unambiguously considered as reality. In the example of a spin of a particle
belonging to the EPR-Bohm pair, the expectation values of spin are defined in
all directions (and equal 0).

 Let us consider the example of the  Schr\"odinger wave of an
electron in a given potential. Our adiabatic measurement allows us to measure
the expectation value of the projection operator on a certain region
of space of   unit volume. We interpret it as a density of the Schr\"odinger
wave. We know that if
we, in a similar adiabatic way,  measure a Gauss flux of the electric
field coming out of this volume we will get the density
times the charge of the electron; if we were able to measure the flux of
the gravitational field we would get the density  times
the mass of the electron; any other property which the electron might have will
manifest itself for any adiabatic coupling as it is spread in the electron
cloud.  Do we want to explain all these facts as a property of the potential
that is  responsible for protecting the Schr\"odinger wave? We may have a
continuum of different potential which all have the property of having
eigenstates with the same density in the chosen volume. There is no
contradiction in calling it just the property of the potentials, but we
believe that there is no contradiction in calling it the density of the
Schr\"odinger
wave either, and the latter choice gives us very powerful intuition for
analysis of various interactions and measurements, interactions which do not
change the Schr\"odinger wave significantly.

Our measurements are not ``measurements'' defined in a standard approach to
quantum theory, i.e.
experiments which specify which eigenvalue the measured variable has. It
seems our usage of  the word ``measurement'' was the  the main reason for
the confusion
generated by our work. We may follow the advice of Bell [9] to abandon the
 the word ``measurement'', and to call our procedure  ``observation''. We can
observe
expectation values of operators, we can observe the density and the current
of the Schr\"odinger wave. We can ``see'' in some sense the Schr\"odinger wave.
This
leads us to believe that it has physical reality.
We hope that recent analysis of possible realization of our ideas in a real
laboratory [10, 11] will soon be implemented and this will serve as evidence
of the  fruitfulness of our idea.

 This research was supported in part by  grant 425/92-1 of the
Basic
Research Foundation (administered by the Israel Academy of Sciences and
Humanities) and by grant PHY 8807812 of the National Science
Foundation.
\vskip 0.8cm
\noindent
\vskip 1cm
{\bf  References}
\vskip 0.32cm

\noindent
1~~ Y. Aharonov and L. Vaidman, {\it Phys. Lett.} {\bf A178}, 38
(1993).\hfill \break
2~~ Y. Aharonov, J. Anandan, and L. Vaidman, {\it Phys. Rev.} {\bf
 A 47}, 4616 (1993).\hfill \break
3~~ C. Rovelli, {\it Phys. Rev.} {\bf A}, to be published. \hfill  \break
4~~ W. G. Unruh, {\it Phys. Rev.} {\bf A 50}, 883 (1994). \hfill  \break
5~~ D. Freedman, {\it Science}, {\bf  259}, 1542 (1993).\hfill  \break
6~~ J. Schwinger , {\it Science}, {\bf  262}, 826 (1993).\hfill  \break
7~~ J. Samuel and R. Nityananda, e-board: gr-qc/9404051.\hfill  \break
%4~~ P. Ghose and D. Home.\hfill \break
8~~ Y.Aharonov
and L. Vaidman, {\it Phys.  Rev.}   {\bf A 41}, 11 (1990).\hfill  \break
9~~  J. Bell, ``Against ``Measurement'', in {\it Sixty-Two Years of
Uncertainty}, A.I. Miller, Ed., Plenum Press, New-York, pp. 17-32 (1990).
\hfill  \break
10~ J. Anandan, {\it  Found. Phys. Lett.} {\bf 6}, 503 (1993).\hfill  \break
11~ S. Nussinov, ``Realistic Experiments for Measuring the Wave Function
of a Single Particle'', UCLA preprint UCLA/94/TEP/2.
\end